\documentclass[sigconf]{acmart}

\AtBeginDocument{%
  }

\setcopyright{acmlicensed}
\copyrightyear{2018}
\acmYear{2018}
\acmDOI{XXXXXXX.XXXXXXX}

\acmConference[IWSPA '25]{11th ACM International Workshop on Security and Privacy Analytics}{June 6, 2025}{Pittsburgh, PA, USA}

\acmISBN{978-1-4503-XXXX-X/18/06}

\begin{document}

\title{A Framework for Cryptographic Verifiability of End-to-End AI Pipelines}


\author{Kar Balan}
\email{k.balan@surrey.ac.uk}
\orcid{0009-0004-1296-9254}
\affiliation{%
  \institution{DECaDE Centre for the Decentralized Digital Economy, University of Surrey}
  \country{Guildford, UK}}

\author{Robert Learney}
\email{robert.learney@digicatapult.org.uk}
\orcid{0000-0003-2870-6120}
\affiliation{%
  \institution{Digital Catapult}
  \city{London}
  \country{UK}
}

\author{Tim Wood}
\email{tim.wood@digicatapult.org.uk}
\orcid{0000-0003-1082-4321}
\affiliation{%
  \institution{Digital Catapult}
  \city{London}
  \country{UK}
}

\renewcommand{\shortauthors}{Balan et al.}

\begin{teaserfigure}
    \centering
    \includegraphics[width=\textwidth]{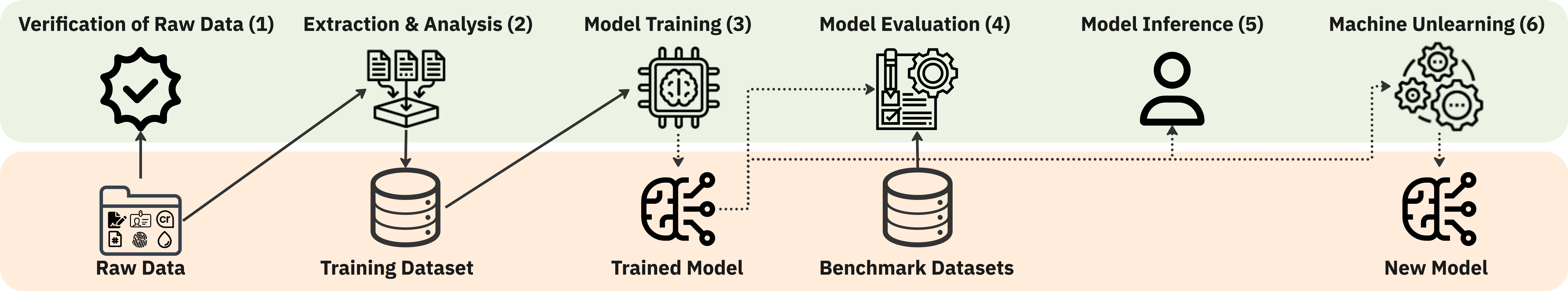}
    \caption{\small Framework for end-to-end verifiable AI pipelines: top-row icons represent Zero-Knowledge Proofs for process integrity; bottom-row icons represent Cryptographic Commitments to data objects. Solid/dashed arrows indicate data/model inputs \& outputs of ZKPs, respectively. }
    \label{teaser}
\end{teaserfigure}

\begin{abstract}
The increasing integration of Artificial Intelligence across multiple industry sectors necessitates robust mechanisms for ensuring transparency, trust, and auditability of its development and deployment. This topic is particularly important in light of recent calls in various jurisdictions to introduce regulation and legislation on AI safety. In this paper, we propose a framework for complete verifiable AI pipelines, identifying key components and analyzing existing cryptographic approaches that contribute to verifiability across different stages of the AI lifecycle, from data sourcing to training, inference, and unlearning. This framework could be used to combat misinformation by providing cryptographic proofs alongside AI-generated assets to allow downstream verification of their provenance and correctness. Our findings underscore the importance of ongoing research to develop cryptographic tools that are not only efficient for isolated AI processes, but that are efficiently `linkable' across different processes within the AI pipeline, to support the development of end-to-end verifiable AI technologies.

\end{abstract}



\received{20 February 2025}
\received[revised]{12 March 2025}
\received[accepted]{5 June 2025}

\maketitle

\section{Introduction}
The rapid adoption of Artificial Intelligence (AI) across diverse sectors has raised significant concerns around the transparency, trust, and auditability of such systems. As AI models become increasingly embedded in critical decision-making, the inability to verify the authenticity, provenance, and operational behaviour of these models poses critical challenges. The increasing use of AI value chains---where data, models, and results are exchanged across various stakeholders \cite{luthi2020distributedledgerprovenancetracking}---creates a need for robust mechanisms and pipelines that ensure trust through verifiability. 

A key scenario illustrating these challenges is that of outsourced model training, where it is important to verify that models are trained using the specified parameters, processes, and datasets \cite{dong2021veridlintegrityverificationoutsourced}. Another challenge arises when different versions of a model are served via opaque APIs: users have no means to verify which model is being served, and may be using one that is cheaper for the server to run rather than the one they paid for \cite{cryptoeprint:2021/730}.  It may not be immediately easy to distinguish between model versions: model quality (e.g. accuracy) may be hard to detect manually. 

These challenges are becoming increasingly relevant in the present context of evolving regulatory initiatives, most notably the European Union’s AI Act \cite{eu_ai_act_2024}, which demands increased transparency and auditability of AI systems, as well as guarantees related to model performance \cite{south2024verifiableevaluationsmachinelearning}. These regulations often mandate third-party audits and the comprehensive review of technical documentation. At the same time, the AI industry is focused on safeguarding the privacy of their models and sensitive data. This trade-off between regulatory demands for openness and the industry’s desire for privacy motivates developing technical solutions to enable trustless audits and verifiability of AI systems without compromising the privacy of underlying data, processes and models \cite{waiwitlikhit2024trustlessauditsrevealingdata}. Cryptographic verifiability offers a high level of automatable assurance and provides a way to achieve strong guarantees of the correctness and integrity of various aspects of AI verifiability:
\begin{itemize}
    \item Digital signatures and cryptographic commitments are used in the C2PA framework (discussed in \S\ref{sec:c2pa}) to provide assurance of media provenance. Cryptographic key material can be held in tamper-proof hardware (e.g. chips within digital cameras) to increase the confidence that a signed image was produced by a particular hardware device.
    \item Zero-knowledge proofs (ZKPs) in various forms allow proving various model properties, including correctness of model training, data provenance and integrity, or execution of model inference, to third parties without revealing inputs they may wish to keep private (such as model weights). 
    \item Federated Learning (FL) allows models to be trained without individual entities revealing their data to a third party. This was notably used by Google to improve autocorrect on Android keyboards while maintaining privacy of individual user inputs \cite{googlefederatedlearning}.
\end{itemize}

The potential for an end-to-end verifiable AI pipeline is only possible through recent advancements in cryptography and verifiable computation due to the high computational complexity of AI training and inference. However, as yet, there are no implementations of this fully verifiable pipeline, although work towards this goal has begun, e.g. \cite{waiwitlikhit2024trustlessauditsrevealingdata}. In this paper, we propose a framework for achieving end-to-end verifiability in AI pipelines. Specifically, we provide the following contributions:

\begin{enumerate}
    \item \textbf{A framework for end-to-end verifiable AI pipelines} (\S\ref{sec:review}): We introduce and outline a framework for verifiability across the AI pipeline, defining the key components required. We review existing solutions and techniques for verifiabality in different parts the AI pipeline---from data provenance and model training, to model evaluation and machine unlearning, and map these approaches onto our framework.
    \item \textbf{Gap analysis} (\S\ref{sec:gaps}): We conduct an analysis to identify the gaps and challenges in achieving full end-to-end verifiability within the AI pipeline. Building on the mapping of existing techniques to our framework, we also propose (at a high level) a generic method for connecting parts of the pipeline together and highlight critical areas where current techniques fall short of achieving full end-to-end verifiability.

\end{enumerate}

We hope that our framework, review of current solutions, and gap analysis motivate further research to develop the necessary tools and technologies to secure the entire AI pipeline. 

\section{Preliminaries}
In this section we give an overview of the cryptographic tools used for AI pipeline verification, which form the technical basis for many of the components in our proposed framework. We also discuss existing methodologies for using these tools in a standardised way. 

\subsection{Cryptographic Tools}
Most cryptographic protocols discussed in this paper involve the use of many cryptographic primitives.  In this subsection, we describe three fundamental primitives: digital signatures, cryptographic commitments, and Zero-Knowledge Proofs (ZKPs).  We refer the reader to the literature for technical definitions.

\subsubsection{Digital signatures}
Digital signatures are ubiquitous in digital infrastructure, notably in X.509 certificates.  They are used to ensure data is from an authentic source and has not been tampered with in transit \cite{rabin1979digitalized}.  There are many security definitions for digital signature schemes. A common security goal is Existential Unforgeability under Chosen Message Attack (EUF-CMA): given a number of message/signature pairs for a fixed public key, it should be computationally infeasible for an adversary to construct a signature for a new message that verifies correctly.

\subsubsection{Cryptographic Commitments}
A cryptographic commitment scheme is a mechanism for `committing' to a message (e.g. a statement or dataset) without revealing it.  A party will create a cryptographic commitment to send to another party, or to post publicly, so that they can `open' it later and reveal the message.  This allows them to declare they know certain information before some event occurs, or to prove their message is not dependent on some external information that has not yet been revealed.  Commitments can also be used in ZKPs (discussed later) so that the information is committed to but need not be explicitly revealed later.

Security in this context means (informally) that the following two properties hold:
\begin{itemize}
\item (Hiding) The commitment reveals nothing about the information committed to.
\item (Binding) The commitment cannot be opened to a different value.
\end{itemize}


Cryptographic commitments are useful in the context of AI pipelines because they allow service providers to commit to information about their model (such as the set of training data) without revealing it directly.  These public commitments can then be verified by a regulator  (where the service providers opens the commitment and shows that they are behaving honestly), or they can be used in ZKPs to allow users to check that outputs they receive are consistent with the public model commitments.  This provides a mechanism for service providers to prove they are serving a particular model to a client (instead of, say, a less powerful model that is cheaper to run) even if the client only has API access. 

\subsubsection{Zero-Knowledge Proofs (ZKPs)}\label{sec:prelims_zk}
We give a high-level summary of ZKPs in this section.  The reader is referred to \cite{ernstberger2024you,bitansky2013succinct} for a formal treatment.

A ZKP is an interactive protocol between a prover and a verifier in which the prover tries to convince the verifier that some statement is true while revealing no information other than this fact.  The protocol must satisfy three properties:
\begin{itemize}
\item (Completeness) An honest prover must be able to convince an honest verifier that a true statement is true.
\item (Soundness) No prover can convince an honest verifier that a false statement is true.
\item (Zero-Knowledge) The transcript between prover and verifier reveals nothing more than that the statement is true.
\end{itemize}
These properties should hold unconditionally, i.e. against a compu\-tation\-ally-unbounded adversary.  A proof where soundness only holds against a computationally-bounded adversary is called an \emph{argument}.  In this work, we use ZKP as an umbrella term that includes zero-knowledge arguments.

A ZKP \emph{of knowledge} is a proof that the prover knows some information.  Intuitively, proving that a statement is true does not necessarily mean the prover knows an answer (`witness') that explains why it is true: for example, a prover can trivially prove that the hash function SHA256 has collisions on the set of 257-bit inputs, but that does not prove knowledge of a specific collision.  A ZKP is called \emph{Non-Interactive Zero-Knowledge} (NIZK) if the prover does not require input from the verifier to generate a proof the verifier still finds convincing.  This is useful where there may be many potential verifiers since the ZKP is essentially `one-way'.

Other common ZKP qualifiers are as follows.  \emph{Succinctness} and \emph{scalability} refer to the efficiency of generating and verifying proofs (the technical details can be found in \cite{DBLP:journals/iacr/Ben-SassonBHR18}).  \emph{Transparency} means no trusted setup is required for the proof system.  Non-transparent schemes require some trusted third party (or additional tools such as Multi-Party Computation (MPC)) to generate unbiased public data that forms the basis of security for the ZKP system.

Proofs that do not satisfy the zero-knowledge property can be useful for demonstrating that a (public) computation was performed honestly without having to re-compute it (which may be costly) \cite{goldwasser2011delegation}.  Well-known proof systems include Succinct Non-interactive Arguments (SNArgs) \cite{micali2000computationally}, Succinct Non-interactive Arguments of Knowledge (SNARKs) \cite{goldwasser2011delegation} and Succinct, Transparent Arguments of Knowledge (STARKs) \cite{DBLP:journals/iacr/Ben-SassonBHR18}.  They are prefixed with \emph{zk-} when they also satisfy the zero-knowledge property (e.g. zk-STARK).


\subsection{Authentication Standards}

\subsubsection{Cryptographic Infrastructure}
Crypto\-graphic standards are important for ensuring that different low-level implementations are interoperable. Digital signature schemes have been standardised in many places, e.g. by NIST and ISO.  By contrast, ZKPs have only recently seen adoption in industry, so there is currently no widely adopted standard, although IETF drafts have been written to attempt to standardise approaches \cite{rfc8235}.  ZKProof (\url{https://zkproof.org/}) exists to support standardisation of ZKPs. 

Standards for cryptographic `infrastructure' go beyond the low-level primitives to provide a common way of communicating cryptographic artefacts regardless of the algorithm specifics.  For example, ECDSA and RSA digital signatures can be used interchangeably in X.509 Public Key Infrastructure (PKI).  To the authors' knowledge, there are currently no standards for infrastructure supporting the distribution of ZKPs except those confined within specific ecosystems (such as in certain blockchains).  This is partly because ZKPs are continually being improved through research efforts, providing a moving target that makes standardisation challenging.

\subsubsection{Content Credentials Metadata}\label{sec:c2pa}
The Coalition for Content Provenance and Authenticity is the cross-industry standards group that developed C2PA - an open specification for encoding provenance metadata within media and other binary assets \cite{c2pa2024}.
C2PA encodes data provenance through a data structure called a `manifest' which contains information, referred to as `assertions', detailing the asset’s origin, including who created it, how, where, when, as well as the source assets (`ingredients') used. Should the ingredient assets also bear C2PA manifests, a graph is created, containing the provenance information of all ingredients. Multiple assertions are grouped within a claim and cryptographically signed with keys authenticated using standard PKI. C2PA securely creates bindings between the manifest and asset. Hard bindings rely on cryptographic hashes to ensure that the manifest belongs to a specific asset and that the asset has not been altered. For images, this involves hashing pixel data or specific file structures, while for non-media assets, it involves hashing raw data or defined byte ranges. Soft bindings use techniques such as content fingerprints or invisible watermarks to identify assets even after transformations or format changes which invalidate hard bindings. The ontology of assertions is flexible and can be extended by the community, containing flags such as 
$\texttt{data\_mining}$, $\texttt{ai\_inference}$, and $\texttt{ai\_training}$, which can empower creatives to specify training consent for their digital assets.

Content credentials are increasingly being adopted across various industries. C2PA manifests are now automatically generated and embedded by several camera manufacturers \cite{sony2024} and creative tools \cite{camerabits2024}; generative models, such as Adobe Firefly and OpenAI DALL-E, are automatically creating C2PA manifests for their generated assets, marking them as synthetically generated \cite{openai2024}.

\section{Framework for End-to-End Verifiable AI Pipelines}

In this work, we propose our \emph{framework for end-to-end verifiable AI pipelines}.  This involves cryptographically verifying that each process in the pipeline has been executed honestly and that the inputs to each process are indeed the outputs of the previous process, shown in fig. \ref{teaser} and summarised below:


\textbf{1. Verification of raw dataset}: Verifying that a corpus of raw inputs, such as images and text, that has been compiled as the initial dataset, contains authentic assets (e.g. by checking digital signatures). Additional properties, such as dataset bias or representativeness, may also be assessed and verified during this phase through Zero-Knowledge Proofs of Fairness (ZKPoFs).

\textbf{2. Extraction \& Analysis}: Collating, processing, and analysing data from the raw dataset to compile the final training dataset. Transformations such as normalisation or conversion to tensor format are applied at this stage to prepare for training. Current pipelines fail to provide full verifiability during this phase.

\textbf{3. Model Training}: Training the model using the training dataset from above as input. Recent advances have made it feasible to apply cryptographic proofs to training, particularly addressing privacy concerns during audits. Zero-Knowledge Proofs of Training (ZKPoTs) enable stakeholders to verify that training was conducted correctly using a specific dataset and algorithm, providing confidence in the model's development process.

\textbf{4. Model Evaluation}: Assessing whether a model satisfies its design requirements. Proofs can be generated to confirm model performance on benchmark datasets, relevant for regulatory contexts like the AI Act. These proofs allow for the private aggregation of model performance metrics, ensuring verifiable evaluations without revealing sensitive data.  ZKPoFs may be used here.

\textbf{5. Model Inference}: Deploying and serving the model to users either directly (e.g. publishing model weights) or via API access to the model running on a server. Zero-Knowledge Proofs of Inference (ZKPoIs) enable a party to prove that an AI model has made a correct inference using a specific, previously committed to model, without disclosing the model weights.

\textbf{6. Machine Unlearning}: Verifying removal of specific training data points from AI models without retraining from scratch or disclosing sensitive information.  Zero-Knowledge Proofs of Unlearning (ZKPoUs) enable proving this without revealing what data was removed (although it is often required that this be known, so the zero-knowledge property may not be needed).  This is crucial for compliance with data privacy laws like the GDPR.

Recent focus on AI verifiability has been on providing assurance over isolated parts of this pipeline: e.g., ZKPoTs, discussed in \S\ref{ZKPoT}, focus solely on proving that the training process was executed honestly given a set of inputs. However, to the authors' knowledge no work has yet achieved a full end-to-end verifiable pipeline.  Such a pipeline would enable an end user who obtains some model inference to verify cryptographically that the output they received was honestly computed using a specific model, and that this model was trained according to a specific algorithm on a specific dataset.  Importantly, cryptographic commitments can be used to commit to various aspects of the model (e.g. model weights) without directly revealing them, supporting privacy while enabling auditability.

Not all use cases demand a fully verifiable AI pipeline, such as where generative AI is used in informal contexts. In some use cases, it may be important for users to be able to verify one or two steps in the pipeline. In security-critical or commercial settings, users may need assurance both that the model was trained honestly from a set of (possibly public) inputs, and that inference was honestly evaluated by the model on their input.  A high level of assurance can be achieved by `connecting' proofs for different parts of the AI pipeline cryptographically. This challenge motivates the study and development of a comprehensive framework for an end-to-end verifiable AI pipeline, which is the central focus of this paper.

We see significant potential for regulatory oversight and compliance frameworks to leverage verifiable AI pipelines. Regulatory authorities, such as those under the AI Act, could play a key role in verifying cryptographic proofs across the pipeline to ensure compliance with legal and ethical standards. Integrating our verifiability framework into regulatory standards could help align AI innovation with public interest and legal safeguards.

In \S\ref{sec:review}, we map existing techniques for verifying components the AI pipeline onto our proposed framework. In \S\ref{sec:gaps}, we discuss the key gaps and challenges preventing full end-to-end verifiability.

\section{Mapping Current Tooling to the Framework}\label{sec:review}

We now explore the key technologies and tools currently available for achieving verifiability across the end-to-end AI pipeline. We review existing solutions and techniques that contribute to verifying various stages, from data provenance and model training to inference and unlearning and map these tools onto the components of the proposed framework, highlighting the current state of the art in AI pipeline verifiability.

\subsection{Data and Transformation Verification Mechanisms (Framework Steps 1 \& 2)}
\label{provenance}
In the AI pipeline, data provenance refers to the detailed record of the origin of data and transformations made on it before being ingested by training algorithms. It is generally achieved using metadata, fingerprinting, and watermarking, which have been described as the `three pillars of data provenance' \cite{authbeyond}.

Verifiable data provenance has become critical in the modern era as digital content manipulation and generation technologies advance. It is particularly important in the context of AI training to ensure AI models are unbiased and robust. Without verifiable provenance, there is a risk of training AI systems on manipulated, biased, erroneous, or copyrighted data without creator consent, all of which undermine the integrity of AI systems. This can accentuate real-world problems and lead to serious consequences such as privacy breaches \cite{carlini2021extractingtrainingdatalarge}, the spread of misinformation, or the creation of non-consensual imagery \cite{Longpre2024Data}.

One key technology aimed at verifying the authenticity of digital images and their provenance is the C2PA standard described in \S\ref{sec:c2pa}, which allows cameras to digitally sign each photo alongside important metadata such as time and location. 
However, this approach has some limitations. If an image undergoes any form of editing—such as cropping, resizing, or applying filters—the original camera signature no longer corresponds to the edited image and cannot be used to verify the authenticity of the modified version. C2PA-enabled editing software can re-sign the edited image, but this presumes trust in the editing software. If a malicious actor compromises this software, they could forge signatures on any image, undermining the trust in the system. C2PA supports integration with Self-Sovereign Identity (SSI) and Verifiable Credentials (VCs), which are mechanisms for establishing identity in a decentralised and secure manner. By leveraging decentralised identifiers (DIDs), VCs enable entities involved in creating or modifying digital assets to issue cryptographically verifiable claims about their identity. These claims, attached to the content, provide assurances about the origin and authenticity of the asset.


Balan et al. propose DECORAIT \cite{decorait}, a decentralised tamper-evident registry for storing C2PA manifests using blockchain technology. DECORAIT enables the retrieval of provenance metadata for images even if the original metadata has been stripped or if the images have undergone visual editing or transformations using robust image fingerprinting. The system ensures that provenance information remains accessible despite metadata removal, eliminating the need for trust in a centrally-managed manifest registry.

ZKPs allow images to be verifiably transformed without revealing the original image or requiring trust in editing software. PhotoProof \cite{photoproof} uses zk-SNARKs to prove that an image which has undergone transformations such as cropping, flipping, transpose, rotation and brightness adjustment, corresponds to an original image that is authenticated with a digital signature computed in secure hardware. However, its key and proof generation times (367 and 306 seconds for 128x128 images) make it impractical for real-world deployment. Kang et al. propose ZK-IMG \cite{kang2022zkimgattestedimageszeroknowledge}, a library that achieves a 100x speed-up over PhotoProof, handling HD image transformations efficiently on commodity hardware. 

Ko et al. \cite{effificentredacting} introduce an efficient verifiable image redacting scheme using SNARKS 
and supports modifying UHD images with proving and verification times under one second.

Datta et al. \cite{veritas} propose VerITAS - a system combining zk-SNARKs with the C2PA framework to verify that permissible image transformations, such as cropping or resizing are correctly applied to C2PA-signed authentic images \cite{AP_TellingTheStory}. These proofs integrate with the C2PA standard, ensuring that the image’s provenance metadata remains verifiable even after transformations, thus preserving authenticity without exposing sensitive content. 

While research and technological progress have been predominantly concentrated on image transformation verification, similar techniques can be applied to audio data. Kang et al. \cite{Kang2023} propose a method using attested microphones and zk-SNARKs to ensure the authenticity of audio recordings and transformations. Although attested microphones, which cryptographically sign audio to verify its source, are not yet commercially available, Kang et al. simulate this functionality using cryptographic keystores (`wallets') on the Ethereum (\url{https://ethereum.org/}) blockchain to sign audio recordings. zk-SNARKs are then employed to validate the authenticity of the audio edits while preserving the privacy of the original content.

In a recent talk, Asaf Shen, Senior Director at Qualcomm, emphasised the need for attested sensors across all types of data to enhance trust and verifiability for various data formats \cite{Shen2024}.

Towards verifiability of unbiased (`fair') training, OATH \cite{oath} provides a framework for allowing users to confirm inference outputs come from models trained fairly with respect to various `demographic variables' (e.g. sex, race, location).  Similarly, Confidential-PROFITT proves that decision trees were trained fairly, starting from commitments to trained models \cite{shamsabadi2023confidentialprofitt}.

\subsection{Zero-Knowledge Proof of Training (Framework Step 3)}
\label{ZKPoT}

Zero-Knowledge Proofs of Training (ZKPoTs) enable a model trainer to demonstrate that a machine learning (ML) model was trained according to some public algorithm. Traditionally, ZKPoTs have been considered computationally prohibitive due to the complexity and scale of modern ML models, and until very recently it was deemed impractical to apply ZKP systems to large-scale neural network training \cite{zkdnn}. However, recent advances in ZKP techniques have made ZKPoT feasible for useful, albeit limited, models.

There are various options regarding privacy in ZKPoTs: for example, training data could be kept as a private input with corresponding cryptographic commitments as a public input so that the data itself need not be revealed when verifying the proof. This could improve privacy during audits: model providers could respond to audit requests by privately computing any function of the dataset (or model) and releasing its output alongside another zero-knowledge proof certifying the correct execution of the function \cite{waiwitlikhit2024trustlessauditsrevealingdata}.

Similarly, the ZKPoT could assert that some public commitment is indeed a cryptographic commitment to a model trained on particular data according to a specific algorithm, without directly revealing the model weights. ZKPoTs are therefore relevant in scenarios where ML models are trained by a party that needs to provide assurances about their training process in regulatory contexts, such as the EU AI Act, to ensure the integrity and correctness of the training process while maintaining data and model privacy.  Notably, ZKPoTs do not hide the training process (which is often a trade secret) since it is essentially encoded in the (public) proof and/or verification algorithm. 

There are still relatively few publications specifically addressing proofs of training; we review them in this section. We note that the majority of early ZKPoT papers focused on simple models, such as logistic regression and decision trees, whereas real-world applications typically demand more sophisticated models like Convolutional Neural Networks (CNNs) and Deep Neural Networks (DNNs). Our focus is therefore on works that extend ZKP capabilities to more complex architectures. We also review works on ZKPs in fine tuning (\S\ref{finetuning}) and federated learning (\S\ref{FL}), which are both part of the training process in the AI pipeline.

Xing et al. \cite{xing2023zeroknowledgeproofmeetsmachine} explore verifiable execution, training, and inference in machine learning, including methods for outsourced machine learning. However, many approaches are not zero-knowledge, as they assume the inputs and model are fully visible to both the prover and verifier, making them unsuitable for end-to-end pipelines where privacy is necessary. Their survey identifies two key challenges in this research area: generalisability and efficiency. Generalisability refers to the fact that, in many ZKP systems, floating-point computations in ML algorithms can only be simulated via arithmetic in large finite fields, which is often either costly or induces precision loss. Efficiency challenges arise from the fact that machine learning tasks are inherently computationally expensive, and existing ZKP systems are typically slow and memory-intensive even for much simpler computations. Indeed, even the most efficient generic ZKP systems are insufficient for proving correctness of circuits as large as the training process of a ML model, without significant optimisation \cite{zkdnn}. In order to address these challenges, researchers have explored various optimisation strategies.  The survey categorises the approaches into three groups: optimising computation processes to reduce complexity, using polynomial approximations (e.g., for sigmoid and ReLU) to minimise verification costs at the expense of some precision, and directly representing activation functions through more complex operations.


zkMLaaS \cite{zkmlaas} and VeriML \cite{veriml} are privacy-preserving frameworks for Machine Learning as a Service (MLaaS) that provide probabilistic rather than full verification of model training through challenge-response protocols and random sampling to generate proofs for specific training epochs and iterations. zkMLaaS maintains model accuracy with only 1\%–4\% degradation resulting from data quantisation and non-linear function approximation. VeriML ensures correctness, fair payment, and trustworthy resource accounting via blockchain-based smart contracts. VeriML supports six ML algorithms and incorporates optimisations such as the Remez approximation and quadratic ReLU approximations. For the CNN model trained on the MNIST dataset with a batch size of 128, VeriML reports a proof generation time of 16.7 s and a verification time of 1.231 s, compared to a native execution time of 1.523 s.


zkDL \cite{zkdl} is a ZKP system for deep learning with two main contributions: 1) zkReLU---a specialised ZKP for the ReLU activation and backpropagation which captures non-arithmetic relationships by leveraging auxiliary inputs to reconstruct intermediate tensors; 2) FAC4DNN---converts deep learning into an arithmetic circuit, efficiently aggregating proofs across layers and steps. Using GPU parallelisation via CUDA, zkDL significantly improves proof times, generating proofs in under a second per batch update for an 8-layer, 10M-parameter network with batch size 64.

ZKAUDIT \cite{waiwitlikhit2024trustlessauditsrevealingdata} is a ZKP-based cryptographic protocol for trustless audits of deep neural networks, preserving the privacy of model weights and training data. The protocol has two phases: ZKAUDIT-T, for generating ZKPoTs, and ZKAUDIT-I, for proving inference results (see \S\ref{ZKPoI}) or other audit functions. The paper makes significant contributions in creating efficient ZKPs for stochastic gradient descent on modern neural networks, implementing optimisations like rounded division in finite fields and high-performance softmax operations. 
Audit scenarios explored include copyright audits, censorship detection, and counterfactual audits, with costs of running ZKAUDIT ranging from \$108 for a copyright audit to \$8,456 for a comprehensive counterfactual audit.

Abbaszadeh et al. propose KAIZEN \cite{zkdnn}, a novel approach to ZKPoTs tailored to deep neural networks. It is designed to handle multiple iterations through a recursive structure, allowing the prover to efficiently validate the execution of gradient descent. By leveraging a recursive sumcheck protocol, KAIZEN amortises the proving costs over multiple training iterations, significantly reducing the overhead typically associated with each iteration. 

\subsubsection{Fine-Tuning (Framework Step 3)}
\label{finetuning}
Fine-tuning is a widespread technique in modern machine learning which adapts pre-trained models to specific tasks by further training on task-specific datasets. This enables high performance on specialised domains without the need to retrain entire models from scratch \cite{ibm_finetuning}. Fine-tuning is as critical as the initial training process in the current landscape of AI system development. Open- or closed-source foundational models are increasingly being sold, licensed, or released to third parties for further adaptation. Fine-tuned open-source models have been shown to outperform generalist proprietary models in specific tasks, driving widespread fine-tuning across various industries \cite{zhao2024loraland310finetuned}. This introduces additional layers of accountability, as the parties performing substantial fine-tuning may be considered providers, rather than deployers of AI systems under the EU AI Act \cite{eu_ai_act_2024}. Consequently, it is necessary to extend verifiability capabilities to fine-tuning in order to support regulatory audits and ensure compliance across the entire AI development lifecycle.

Fine-tuning methods can be broadly categorised into full fine-tuning, which involves updating all the parameters of a pre-trained model, and parameter-efficient fine-tuning (PEFT) approaches, which update only a small subset of the model’s parameters while keeping most of the pre-trained weights frozen \cite{zhao2024loraland310finetuned}, presenting distinct computational and practical characteristics relevant to ensuring verifiability. The technical requirements for ZKPs of fine-tuning largely align with those for pre-training. For full fine-tuning, existing ZKPoT schemes can be adapted with minor adjustments to accommodate the new dataset and weights initialisation. However, for PEFT, separate circuits must be designed to validate the specific operations performed, such as updating only a subset of weights.

Garg et al. \cite{garg} introduce the first work to explicitly address the concept of zero-know\-ledge proofs of fine-tuning. The paper does not provide specific technical implementation details of the proposed ZKP system for fine-tuning, but does propose methods to ensure the integrity of inputs from foundational models. For instance, the authors suggest using digital signatures to verify the provenance of the committed inputs. Moreover, the system requires the prover to demonstrate not only that the model was trained honestly but also that the input data is well-formed, proving the prover's knowledge of valid signatures for the committed inputs. 

vTune \cite{zhang2024vtune} enables verifying whether an LLM has been fine-tuned as claimed by embedding `backdoors' into training data. The presence of these backdoors can be detected in the model output, providing statistical evidence of fine-tuning without requiring access to the weights. However, vTune's reliance on backdoors introduces limitations---it does not guarantee granular aspects like specific fine-tuning methods, it is potentially vulnerable to adversarial attacks and shows reduced performance in certain tasks.
\subsubsection{Federated Learning  (Framework Steps 1, 2 \& 3)}
\label{FL}

Federated Learning (FL) is a type of distributed machine learning that enables multiple participants to collaboratively train a model without sharing their raw training data with a central server. Typically, clients train the model locally on their devices and only share model updates (e.g., gradients or weights) with a central aggregator; however, aggregator-free systems have also been proposed \cite{pdfed}. FL significantly enhances training data privacy, however, this introduces additional challenges in verifying the authenticity of client-provided model updates and ensuring the security of the training and aggregation processes. Integrating ZKPs into FL frameworks can provide cryptographic guarantees that model updates are valid---meaning they were correctly trained according to an agreed-upon protocol, which serves as proof of training---without revealing any additional information about the underlying data.

Regulatory frameworks, such as the EU AI Act, apply to FL participants, who may be considered `providers' under certain circumstances, subjecting them to additional legal responsibilities, including ensuring compliance with data protection regulations and maintaining transparency in their processes. Thus, the ability to provide proof and be subject to auditing is crucial for compliance and accountability. Incorporating ZKPs naturally complements this, enabling proof of correctness in a privacy-preserving manner.

FL participants could be required to commit to their data publicly to enable the auditing of model inputs, disincentivising dataset manipulation to poison models \cite{MA2024112115}.  In addition, ZKPs could be used to prove that a client has followed the expected learning algorithm, or that their training data satisfies certain predicates (e.g. numerical values lie within certain bounds), ensuring consistency and fairness across participants. In systems with a central aggregator, ZKPs can also be employed to guarantee that the aggregation of model updates was performed correctly and without tampering \cite{zkfl}. Another relevant use case for ZKPs is proving ownership in FL scenarios where multiple stakeholders contribute to the development of a model. This could play a vital role in managing rights, rewards and responsibilities within collaborative AI ecosystems, helping to maintain trust among participants.

Ruckel et al. \cite{RUCKEL2022108621} provide a proof of concept implementation for a FL framework involving multiple linear regression models, which allows clients to prove that they have truthfully trained their model updates based on commitments to private data. Additionally, the system can incorporate a proof of provenance, ensuring that data originated from a certified source, such as an attested sensor as discussed in \S\ref{provenance}. Differential privacy techniques are also applied to further protect the privacy of local weights.

Wang et al. introduce PTDFL \cite{WANG2023449}, a decentralised, aggregator-free FL framework. Each node conducts local training on its dataset and generates a proof of training to validate the correctness of the computed gradient. This local proof, along with the encrypted gradient ciphertext, is broadcast to the other participating nodes. Each node verifies these proofs before aggregating the gradient ciphertexts, ensuring that only trustworthy model updates contribute to the global model, maintaining the integrity of the aggregation process without relying on a central trusted entity.

martFL \cite{martFL} is an FL framework that facilitates a secure, utility-driven data marketplace for AI models. It addresses several key challenges, including the private assessment of models prior to acquisition using CKKS homomorphic encryption \cite{cheon2017homomorphic}, verifiable model aggregation, and the fair, automatically enforceable distribution of rewards through smart contracts. martFL enables the `data acquirer' (model buyer) to prove in zero-knowledge that it has accurately aggregated local models according to pre-committed weights and ensures that `data providers' (model trainers) can claim rewards that are proportionate to their contributions.

Flais et al. \cite{flaiact} discuss the implications of the EU AI Act for FL, emphasising the requirements for clear stakeholder responsibilities in the development and deployment of AI systems. Although all parties bear some degree of liability under the AI Act, the authors propose a strategic shift to designate the server operator as the primary service provider. One significant challenge arises from the stipulation that any data involved in training processes must be evaluated for potential bias or adversarial information, which poses difficulties for the server, as the training data resides within local nodes, limiting its ability to conduct thorough evaluations. Here, ZKPs can support compliance by providing cryptographic assurances that model updates are derived from previously committed training data, enabling clients to demonstrate the integrity of their contributions without necessarily exposing sensitive information. Additionally, ZKPs can facilitate the verification of certain attributes related to the training data through methods such as verifiable evaluations, as discussed in \S\ref{ZKPoI}, and verifiable provenance, outlined in \S\ref{provenance}. However, as further explored in \S\ref{gap1}, fully verifiable data processing remains an unresolved challenge.

In summary, the integration of ZKPoTs has the potential to open up new possibilities for privacy, auditability, and compliance in both centralised and distributed AI. By leveraging ZKPs, systems can mitigate risks related to data manipulation and model poisoning, thereby enhancing the security and accountability of collaborative AI ecosystems. However, significant limitations exist---notably the inability to keep the model architecture private due to its representation as the ZKP circuit itself. This presents challenges in scenarios where model architecture confidentiality is essential. One possibility for addressing this concern is for regulators to verify proofs privately and issue certificates to service providers instead of proofs being released publicly. Despite this, ZKPs still allow for the privacy of the resulting weights, as well as input data. Additionally, while ZKPoTs ensure honest training by preventing data duplication and manipulation, they fall short of providing comprehensive guarantees of model correctness---there remain opportunities for biases. Research focusing on how ZKPs might be combined with alternative model fairness checking methods to enhance guarantees of correctness and mitigate potential biases would be beneficial.

\subsection{Zero-Knowledge Proof of Inference (Framework Steps 4 and 5)}
\label{ZKPoI}

Zero-Knowledge Proofs of Inference (ZKPoIs) allow one party (the prover) to demonstrate to another party (the verifier) that a specific machine learning model has made a correct inference on a given input, while keeping the input or the model (or both) private. This ensures the privacy of both the data and the model, which can be particularly important in sensitive applications such as healthcare  and security. The ability of ZKPoIs to prove the accuracy of model predictions without exposing the underlying model weights is especially relevant in regulatory contexts like the AI Act, which mandates ensuring model accuracy and robustness. ZKPs can be generated during model inference to produce proofs of the model’s outputs on benchmark datasets, allowing for verifiable attestations that confirm the model’s accuracy and performance \cite{south2024verifiableevaluationsmachinelearning}. Another important use case involves verifying that the correct version of a model is being served, particularly in cases where users pay for API access without visibility of the underlying model. ZKPs further enable trustless verification in scenarios like legal subpoenas, allowing a responder to use an ML model to privately retrieve and verify documents that match a request, without exposing the underlying data or tampering with the results \cite{zkml}. Additionally, as AI models are increasingly involved in decision-making processes, proofs of inference enable establishing a verifiable chain of provenance for decisions made using AI models. Two surveys \cite{modulus2023cost, xing2023zeroknowledgeproofmeetsmachine} provide analyses of various ZKP systems, offering insights into the trade-offs between performance, scalability, and privacy.

The Modulus Labs whitepaper \cite{modulus2023cost} benchmarks multi-layer perceptron (MLP) architectures against a set of ZKP systems for inference, including Groth16 (\url{https://github.com/arkworks-rs/groth16}), Gemini \cite{bootle2022gemini}, halo2 (\url{https://zcash.github.io/halo2}), Plonky2 (\url{https://github.com/0xPolygonZero/plonky2}), Winterfell (\url{https://github.com/facebook/winterfell}) and zkCNN \cite{liu2021zkCNN}. 
Their experiments scale with increasing parameter counts (up to 18M parameters) and network depth (up to 500 layers). The analysis highlights Plonky2 as the best performing in terms of proof generation time, outperforming halo2 by a factor of five in some cases. For 15M parameters, proof generation times vary significantly, with halo2 taking around 275 seconds compared to Plonky2’s 50 seconds. However, Plonky2’s memory consumption is notably higher, with a network of 11M parameters requiring 42GB of memory compared to halo2’s 19GB. As the model size increases to 18M parameters, the memory usage gap narrows, with halo2 consuming 48GB and Plonky2 56GB. zkCNN performs best in handling larger models, achieving top two results across experiments for both proof generation time and memory efficiency, making it a suitable choice for deep learning applications. These findings provide a strong foundation for future work, as most solutions build upon these foundational ZKP systems.

Xing et al. \cite{xing2023zeroknowledgeproofmeetsmachine} review developments in verifiable inference and execution up to June 2023, emphasising the key challenges of generalisability and efficiency, discussed previously in the context of ZKPoTs (\S\ref{ZKPoT}), which apply equally to verifiable inference. 

Kang et al. \cite{kang2022scalingtrustlessdnninference} presented the first ZKP scheme for ImageNet-scale MobileNet v2 using the halo2 framework to create an efficient zk-SNARK circuit, unlike prior work which focused on significantly smaller models. They optimised division verification with custom gates for linear layers and a lookup argument for non-linear layers, while ensuring privacy through hash commitments. 

Building on previous work, Kang et al. \cite{zkml} introduce ZKML, a framework designed to generate zk-SNARKs for a range of modern machine learning models, including large-scale models like GPT-2 and Stable Diffusion, unlike prior schemes, which were limited to CNNs.
The accuracy of models such as VGG16 and ResNet-18 differed by only 0.01\% from their floating-point counterparts, demonstrating minimal performance degradation despite the quantisation and arithmetisation changes required for zk-SNARKs. Kang et al. emphasise the potential for ZKML in enabling trustless audits for machine learning systems, particularly those that rely on private data, however, a notable limitation of ZKML is that it requires the model architecture (but not the weights) to be revealed.

ezkl (\url{https://ezkl.xyz/}) is currently the most mature and comprehensive open-source library for integrating zero-knowledge proofs with ML models, however it is still in its early beta stage, undergoing rapid development and is not yet suitable for use in production environments. Built on the halo2 proving system, it enables verifiable and privacy-preserving inference by automatically converting AI models from ONNX format into zk-SNARK circuits.

South et al. \cite{south2024verifiableevaluationsmachinelearning} extend the concept of verifiable machine learning evaluation by using zk-SNARKs to create proofs of model performance---proving that a model achieves specific performance or fairness metrics on public benchmarks. Similarly to ZKML, the model weights are kept private but the model architecture is public. South et al. propose a `predict, then prove' strategy which allows model inference results to be delivered immediately, while the proof can be generated later. This is useful in contexts where auditability is important, but the proof generation time is less critical. Individual inference proofs can be na{\"i}vely bundled into a single package, which is computationally efficient but exposes inference outputs, presenting privacy risks. Alternatively, an aggregation circuit can be employed to validate each proof and consolidate results, adding a layer of security but increasing computational complexity. Another approach involves constructing a custom halo2 circuit that aggregates only essential metrics, such as accuracy scores or confusion matrices, allowing for private data outputs while sharing verifiable performance metrics. Notably, the nanoGPT model exhibits the highest resource consumption, with a proving time of 46 s, verification time of 2.69 s, and a proof size of 0.7 MB. The experiments indicate that proof times and sizes scale linearly with the number of constraints and multiply-accumulate operations (MACs), especially for complex architectures like transformers. 

FairProof \cite{fairproof} is a system that uses ZKPs to verify the fairness of machine learning models. It consists of two main components: a fairness certification algorithm for fully-connected neural networks, which generates a personalised `fairness certificate' for each data point, and a ZKP-based cryptographic protocol to verify both the proof of inference and the fairness certificate without revealing the model weights, although the model architecture is public. FairProof is tested on individual fairness benchmark datasets and demonstrates practical feasibility. FairProof generates a verifiable fairness certificate in an average of 1.17 minutes per query without parallelisation on an Intel-i9 CPU, with certificate sizes averaging 43.5 KB, and an average verification time of less than 2 seconds.

The importance of ZKPoIs lies in their ability to enhance trust in machine learning systems by providing verifiable correctness without revealing sensitive information, opening the door to broader adoption of machine learning services where transparency and security are key. However, as previously discussed in \S\ref{ZKPoT}, a fundamental limitation remains: while ZKPoI ensures that sensitive information like model weights and input data remain private, the model architecture itself is the ZKP circuit, and is thus publicly accessible. This limitation poses challenges in scenarios where architectural confidentiality is essential. In such cases, a regulator could verify the ZK proofs and issue a compliance certificate to service providers. Achieving both privacy for model structure and practical efficiency in ZKP schemes is crucial for broader industry adoption, making this an important avenue for future research.

\subsection{Zero-Knowledge Proof of Unlearning  (Framework Step 6)}
\label{ZKPoU}

Machine unlearning is defined as the process of removing a specific training data sample and its influence from an already trained machine learning model. A Zero-Knowledge Proof of Unlearning (ZKPoU) addresses the need to verifiably remove data from machine learning models without revealing (for example) current or revised model weights.   This concept is particularly important in the context of data privacy regulations like the General Data Protection Regulation (GDPR)\cite{gdpr2016}, California Consumer Privacy Act (CCPA)\cite{ccpa2018}, and Canada’s Personal Information Protection and Electronic Documents Act (PIPEDA)\cite{pipeda2000}, all of which enforce the `right to be forgotten'. These regulations grant individuals the right to request the removal of their data, which extends to AI models trained using that data. In a recent opinion \cite{edpb2024}, the European Data Protection Board emphasised that supervisory authorities may impose corrective measures, such as model retraining, in cases where unlawful data processing occurs during AI model development. ZKPoUs can be used for the verifiable removal of unlawfully included training data, ensuring compliance with data protection regulations without retraining.

Unlearning methods are broadly categorised into two types: exact and approximate unlearning. Exact unlearning aims to fully remove the influence of specific data points from the model by retraining it from scratch \cite{Xu_2024}. This provides strong guarantees that the data has been removed, but it comes at a significant computational cost, making it more suitable for simpler models. Approximate unlearning focuses on efficiently reducing the influence of certain data points by updating the model’s parameters. Although this method does not completely eliminate the influence of the targeted data, it drastically reduces computational and storage requirements, making it viable for real-world applications \cite{Xu_2024}. Several surveys have explored the topic of machine unlearning, each classifying techniques into distinct taxonomies, e.g. \cite{nguyen2024surveymachineunlearning}.



Eisenhofer et al. \cite{unlearningzkp} present the first ZKP system for verifiable machine learning and unlearning processes. The system supports exact unlearning, as well as two types of approximate unlearning: amnesiac unlearning, which removes the updates to the model’s parameters that were directly computed on the specific data point, and optimisation-based unlearning, in which the loss corresponding to a specific data point is increased and the resulting model parameters are subtracted from the original model. The system uses Spartan \cite{spartan}, a ZKP system which does not require a trusted setup (see \S\ref{sec:prelims_zk}. The system generates three core types of proofs: proof of training, which establishes the integrity of the model prior to any unlearning requests; proofs of unlearning, which ensure compliance with unlearning requests and allow the user to verify that their data is correctly removed from the model; and proofs of non-membership, proving that the specific data point is no longer part of the dataset. One limitation of the system is that users must verify every update to the model to ensure that their data has not been re-added. Users must therefore monitor for updates, which is often impractical in real-world systems. To mitigate this, regulators or trusted third parties could step in to verify unlearning requests on behalf of users and ensure compliance across the system. There is also no mechanism to verify that the updated model has been redeployed and is the one served to users, leaving a gap in the end-to-end verification of the machine learning pipeline. Addressing this gap is necessary for a fully verifiable machine learning system.

Weng et al. present a framework for proof of unlearning in Machine Learning as a Service (MLaaS) settings \cite{unlearningtee}. The system leverages Trusted Execution Environments (TEEs) to maintain the integrity of unlearning processes and training data. The system facilitates efficient retraining by updating only the submodels affected by the unlearning request. While TEEs are typically much more efficient than using ZKPs, they are not immune to security vulnerabilities and can be compromised \cite{teesunsafe}, unlike ZKPs whose security is guaranteed cryptographically.

Gao et al. highlight critical challenges in achieving verifiable machine unlearning, demonstrating that even with sophisticated unlearning mechanisms, privacy attacks can exploit both the original and updated models to infer or reconstruct deleted data \cite{deletion}. This underscores the importance of developing robust unlearning frameworks that not only efficiently manage data removal but also provide strong privacy guarantees.

ZKPoUs could play an important role in ensuring that machine learning models comply with data privacy regulations and the right to be forgotten by providing individuals with the assurance that their data can and has been removed from AI systems. Despite progress in unlearning methods, significant challenges remain, particularly concerning computational efficiency and the need for robust verification mechanisms. The role of regulators or trusted third parties becomes vital to ensure compliance and facilitate the verification of unlearning requests. By addressing these challenges, ZKPoUs could significantly contribute to regulatory compliance an ethical AI practices in an increasingly data-driven world.




\section{Gaps in verifiability of the AI pipeline}\label{sec:gaps}
While existing solutions, as explored in \S\ref{sec:review}, provide  verifiability for individual components across our proposed framework, few of them consider cryptographically connecting the outputs of one step in the AI development pipeline to the inputs of the next. As a result, the pipeline---from raw data through to generated assets or model inference results---remains only partially verifiable. As previously dicussed, a full end-to-end verifiable AI pipeline would allow an end user who obtains some model inference to verify cryptographically that the output they received was honestly computed using a specific model, and that this model was trained according to a specific algorithm on a specific dataset. Reducing reliance on human-driven processes and instead using cryptography to verify correctness is closer to a zero-trust architecture, which has become a focus of secure system design in recent years.

Proving the correctness of the entire pipeline `at once' (i.e. in one monolithic proof) is not a good approach for several reasons.  Firstly, designing such a system is a significant task and possibly requires rework for each model architecture, training data format, etc..  Secondly, re-proving correctness of training for every inference is an unnecessary overhead.  Thirdly, ZKP systems optimised for training will not necessarily be efficient for inference, and vice versa (e.g. where different circuit arithmetisation techniques are used).  Instead, the most natural solution would appear to be for ZKPs at each step in the process to include in their scope the proof of correctness commitments to aspects of the model being trained or evaluated.  This solution could be seen as viewing the AI pipeline as an incrementally-verifiable computation (IVC) \cite{10.1007/978-3-540-78524-8_1}, which is concretely realised in works by Abbaszadeh et al. \cite{zkdnn} and Bowe et al. in halo2, discussed in the previous section.  In practice, this looks like the following: for each process in the pipeline, take all outputs and commitments from the previous process as input, verify the commitments, execute the current process and compute commitments to the new outputs, and generate a proof that the current process and commitment generation were performed honestly.  

The proposed framework for an end-to-end verifiable AI pipeline would further enable comprehensive provenance tracing for both models and their generated outputs. This would help combat misinformation by providing verifiable information regarding training data and attesting to the correctness of the training and inference processes. Ultimately, it would enable users to trace the origins of generated assets and the integrity of the processes that produced them, fostering transparency and trust in AI systems.

The lack of end-to-end AI pipeline verification in the literature appears mainly to be due to the nascency of proving AI processes in zero knowledge and the effort involved in designing ZKPs for the whole pipeline.  Indeed, research papers sometimes provide sketches of how a fuller pipeline could be developed from the solution presented (e.g. descriptions of image input commitments in papers on ZKPoT \cite{zkdnn}).   In the current work, we argue that developments in ZKPs for AI have reached a level of maturity that now warrants investigating the verifiability of the full pipeline, especially in light of emerging regulatory demands. In this section, we outline the main gaps we have identified and discuss how they might be addressed.

\subsection{Gap between raw data and training input data}
\label{gap1}

The first gap identified in the verifiable AI pipeline sits between data in its raw form and data in its processed form, as it enters the training step. While verifiable data provenance (\S\ref{provenance}) is a significant first step towards partially closing this gap and securing the entire AI pipeline, alone it is insufficient. There are two parts to this gap: down-selecting a set of training data from a larger set of raw data through refinement and analysis, and transforming the training data for ingestion by training processes.

The data refinement gap is substantial, encompassing many key data processing stages. Current and upcoming regulatory requirements \cite{eu_ai_act_2024} emphasise the need for robust data governance and data management practices, underlining the importance of documenting data collection processes, the provenance of the data, and the intended purposes for which the data was collected. In many cases, dataset refinement---such as removing NSFW or copyrighted images via automated tools like AI---may be required. Proving the correctness of such processes could itself involve generating and aggregating ZKPoIs of AI-based filtering. This refinement step could ensure compliance with regulatory requirements for dataset preparation and seamlessly integrate with the proposed pipeline.

The second part of this gap covers transformation into suitable formats (e.g., tensors), as well as normalisation and other standard techniques in ML. Each model typically follows a unique `recipe' for these transformations (also known as model-dependent transformations), which is applied to data both at the training step and the inference step. Generating ZKPs for these transformations will sometimes be quite involved: for example, images in training data that have attestations of provenance using C2PA may be in compressed formats (e.g. JPEG, PNG) that need to be decompressed before being used in a training algorithm. Filling this gap could simply involve extending ZKPoTs and ZKPoIs to ingest raw data, internally transforming and proving correctness of transformation as part of the larger proofs.

Xing et al. \cite{xing2023zeroknowledgeproofmeetsmachine} examine this gap from the perspective of multiple independent parties involved in different tasks across the AI pipeline, where it is crucial for these parties to provide proofs of correctness to each other without disclosing additional information. In contrast, our proposed pipeline aims to offer verifiable guarantees even when all processes are conducted by a single actor, enabling auditors or users to independently verify these claims. Xing et al. divide data preprocessing into two sub-tasks: data collection and data refinement. Data collection involves selecting and acquiring data from original data owners based on specific criteria, which often requires querying databases. In the data refinement step, various preprocessing algorithms are applied enhance the features of the data, aiming to improve data quality. Xing et al. conclude that combining these dataset and preprocessing operations with ZKPs to validate correctness is rare. They find that the data collection process often aligns with the concept of Verifiable Databases (VDB), which, while not computationally intensive, face efficiency challenges due to high input-output complexity. Existing VDB solutions frequently use lightweight techniques such as digital signatures, polynomial commitments, or TEEs to improve efficiency. 

Notably, Wang et al. \cite{wang2023ezdpsefficientzeroknowledgemachine} have designed the ezDPS scheme, which is the only work of which the authors are aware that proves the correctness of discrete wavelet transform (DWT) and principal component analysis (PCA) in ML scenarios. Elsewhere, Singh et al. \cite{decaipipelines} address decentralised AI pipelines, where different steps of the AI data processing and training pipeline are carried out by independent actors. The paper identifies the same lack of a unified AI pipeline for verifiability. The authors propose privacy-preserving ZKPs for basic operations such as aggregation, filtering, column sorting, and column inner-joining, but they do not address model-dependent transformations or proofs of training.

\subsection{Gap between ZKPoT and ZKPoI}
While a ZKPoT provides confidence that a model was honestly trained, it only confirms that there exists such a model. ZKPoTs are only useful to an end user receiving model outputs if they can be convinced they are being served the corresponding model and that the inference output was obtained from it.

Similarly, while a ZKPoI provides a user with confidence that their input was honestly evaluated using some model, the proofs only confirm that there exists a model that produces a particular output when evaluated on the user's input.  Additionally, ZKPoIs often take the arithmetised ML model as input, not the model itself.  The user must therefore trust that the arithmetic circuit representation used for the ZKPoI is an `honest' representation of the model.  Linking between ZKPoI and ZKPoT closes this gap in verifiability and gives a user confidence that a service provider is serving the claimed model.

However, closing this gap is not necessarily straightforward.  ZKPoI and ZKPoT schemes often use different, sometimes bespoke proof systems that are not directly compatible, e.g. quantisation is done differently.  This is not an insurmountable challenge, but it has not been a research goal in most ZKPs for AI so far. The ZKAUDIT solution by Waiwitlikhit et al. \cite{waiwitlikhit2024trustlessauditsrevealingdata} does indeed provide ZKPoTs and ZKPoIs that can be linked directly, showing that it is feasible.  Clearly it is also possible to use a black-box approach to connect the proofs, where committed model weights are proven to correspond to model weights used in the ZKPoI through additional ZKPs, but it is not clear that this would be efficient.

\section{Conclusions}\label{sec:conclusions}
In this paper, we proposed a framework for achieving end-to-end verifiability in AI pipelines, leveraging cryptographic tools such as digital signatures, cryptographic commitments, and Zero-Knowledge Proofs (ZKPs) to ensure strong guarantees of model correctness. Our focus was on verifiability of the whole end-to-end AI pipeline, including data extraction and refinement, model training, inference, and unlearning.

In \S\ref{sec:review} we mapped existing solutions and techniques onto the proposed framework, highlighting their potential contributions to enabling compliance with incoming AI regulation (specifically the European Union’s AI Act), and for AI model deployment scenarios where a high level of assurance of the end-to-end AI pipeline is needed. 

In \S\ref{sec:gaps} we identified gaps in the AI pipeline that hinder full end-to-end verifiability, and highlighted research opportunities for bridging these gaps.

Our analysis has led us to following conclusions:
\begin{itemize}
    \item The need for AI verifiability is a driver for developing ZKP standards. Research on ZKPs for AI is reaching a level where it can begin to be used in real-world scenarios to demonstrate compliance. However, there are currently no standard approaches for implementing them. Standards are needed both to enable interoperability of implementations of low-primitives, and also for conveying the cryptographic artefacts using common data formats. As ZKPs provided as metadata to AI model outputs can be removed through malicious action or data corruption, frameworks such as DECORAIT \cite{decorait} may be needed to provide immutable metadata records that bind output to models used to generate them.
    \item Regulators may need to play a critical role in enhancing trust in AI systems by taking on the responsibility of verifying ZKPs. ZKPs do not necessarily hide 
    algorithms such as the training process, so regulators may need to verify  proofs of proprietary models on behalf of users to allow service providers to keep their model architectures private while giving strong cryptographic security guarantees. This approach addresses concerns regarding transparency in AI, particularly in light of the EU AI Act’s requirements for auditability.
    \item There are a number of gaps in the verifiability of the AI pipeline.  These gaps are possible focus areas for future research on full end-to-end verifiability.
\end{itemize}

Through the proposed framework, we hope to motivate the creation of robust, privacy-preserving AI systems that provide verifiable assurance across the entire AI pipeline, ultimately paving the way for more responsible and trustworthy AI deployment.

\begin{acks}
This work was carried out while Kar Balan was an intern at Digital Catapult and was partly supported by UKRI Grant EP/T022485/1.

\end{acks}

\bibliographystyle{ACM-Reference-Format}
\bibliography{sample-manuscript}

\end{document}